\documentclass[aps,prd,reprint,amsmath,amssymb]{revtex4-2}

\usepackage{newtxtext}
\usepackage[varg]{newtxmath}
\usepackage[T1]{fontenc}
\usepackage{graphicx}

\usepackage[stretch=10,shrink=0,final]{microtype}
\usepackage{nowidow}
\let\revtitle\maketitle
\renewcommand{\maketitle}{%
	\revtitle
	\tolerance=1000
	\hyphenpenalty=1000
}

\usepackage{pdfpages}

\makeatletter
\AtBeginDocument{\let\LS@rot\@undefined}
\makeatother

\IfFileExists{SupplementalMaterial.pdf}{%
	\AtEndDocument{%
		\clearpage\onecolumngrid%
		\includepdf[pages={{},-8}]{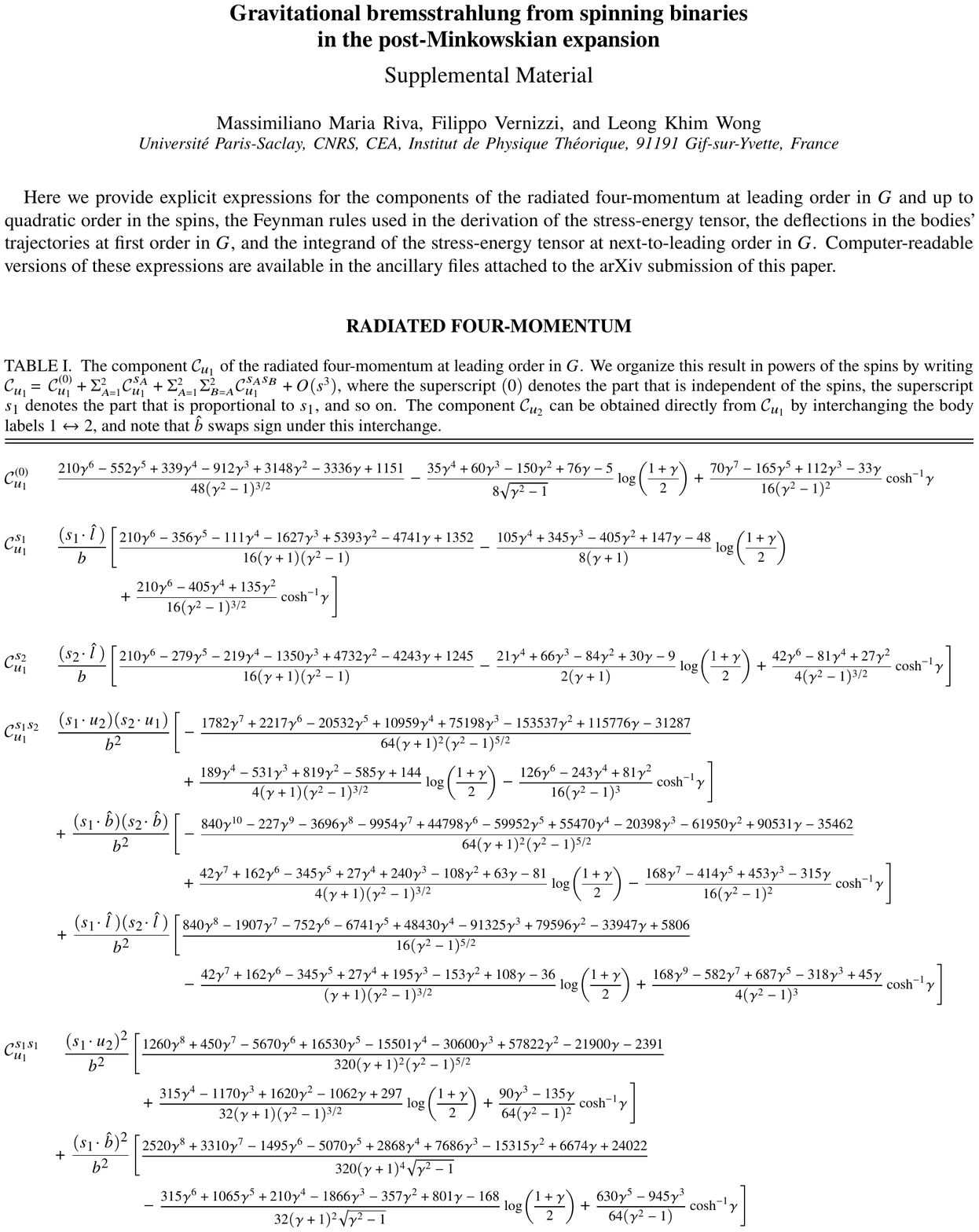}%
		\includepdf[page=9]{SupplementalMaterial.pdf}%
	}%
}{}

\DeclareFontFamily{U}{futm}{}
\DeclareFontShape{U}{futm}{m}{n}{<-> fourier-bb}{}
\DeclareMathAlphabet{\mathbb}{U}{futm}{m}{n}

\DeclareSymbolFont{cmreg}{OT1}{cmr}{m}{n}
\DeclareSymbolFont{cmmath}{OML}{cmm}{m}{i}
\DeclareSymbolFont{cmsymbols}{OMS}{cmsy}{m}{n}
\DeclareSymbolFont{cmlargesymbols}{OMX}{cmex}{m}{n}
\DeclareSymbolFontAlphabet{\mathcal}{cmsymbols}

\DeclareMathSymbol{\partial}{0}{cmmath}{64}
\DeclareMathSymbol{g}{\mathalpha}{cmmath}{103}
\DeclareMathSymbol{\alpha}{0}{cmmath}{11}
\DeclareMathSymbol{\beta}{0}{cmmath}{12}
\DeclareMathSymbol{\eta}{0}{cmmath}{17}
\DeclareMathSymbol{\kappa}{0}{cmmath}{20}
\DeclareMathSymbol{\mu}{0}{cmmath}{22}
\DeclareMathSymbol{\nu}{0}{cmmath}{23}
\DeclareMathSymbol{\rho}{0}{cmmath}{26}
\DeclareMathSymbol{\sigma}{0}{cmmath}{27}
\DeclareMathSymbol{\ell}{0}{cmmath}{96}

\DeclareMathSymbol{\ointop}{\mathop}{cmlargesymbols}{72}
\DeclareMathSymbol{\intop}{\mathop}{cmlargesymbols}{82}
\DeclareMathDelimiter{(}{\mathopen}{cmreg}{40}{cmlargesymbols}{0}
\DeclareMathDelimiter{)}{\mathclose}{cmreg}{41}{cmlargesymbols}{1}

\DeclareMathDelimiter{(}{\mathopen}{cmreg}{40}{cmlargesymbols}{0}
\DeclareMathDelimiter{)}{\mathclose}{cmreg}{41}{cmlargesymbols}{1}
\DeclareMathDelimiter{[}{\mathopen}{cmreg}{91}{cmlargesymbols}{2}
\DeclareMathDelimiter{]}{\mathclose}{cmreg}{93}{cmlargesymbols}{3}

\usepackage{titlesec}

\newcommand{\mysectionnumbering}{\thesection.~}

\titleformat{\section}{\bfseries\center\uppercase}{\mysectionnumbering}{0em}{}
\titleformat{\subsection}{\bfseries\center}{}{0.1em}{\thesubsection.~}
\titleformat{\subsubsection}{\bfseries\itshape\center}{}{0.1em}{\thesubsubsection.~}

\titlespacing{\section}{0pt}{1.7em plus 0.9em minus 0.9em}{0.7em plus 0.2em minus 0em}
\titlespacing{\subsection}{0pt}{1.5em plus 0.1em minus 0.1em}{0.5em}
\titlespacing{\subsubsection}{0pt}{1.5em plus 0.1em minus 0.1em}{0.5em}

\titleformat{\paragraph}[runin]{\itshape}{}{0em}{}[.---\,]
\titlespacing{\paragraph}{\the\parindent}{0em}{0em}

\usepackage{xcolor}
\definecolor{revblue}{HTML}{2d3092}
\colorlet{blue}{revblue} 

\let\revcite\cite
\renewcommand\cite[1]{\mbox{\color{blue}\revcite{#1}}}

\let\reveqref\eqref
\renewcommand\eqref[1]{{\color{blue}\reveqref{#1}}}

\IfFileExists{apsrev-custom.bst}{\bibliographystyle{apsrev-custom}}{}

\newlength\mycitespacing
\setcitestyle{citesep={,\kern-\mycitespacing}}

\usepackage[
	hyperfootnotes=false,
	colorlinks=true,
	allcolors=blue,
	breaklinks=true]
{hyperref}
\newcommand{\X}{X}
\newcommand{\avg}[1]{{\langle #1 \rangle}}
\newcommand{\dx}{\text{d}}
\newcommand{\U}{\mathcal{U}}
\newcommand{\Sp}{\mathcal{S}}
\newcommand{\deltabar}{\ensuremath{\text{\smash{\raisebox{0.6ex}{--}}\kern-1.1ex$\delta$}}}
\newcommand{\urho}{\underline{\rho}}
\newcommand{\dualu}{\check{u}}
\newcommand{\lhat}{\hat{l}}
\newcommand{\bhat}{\hat{b}}

\begin{document}
\title{Gravitational bremsstrahlung from spinning binaries\protect\\in the post-Minkowskian expansion}

\author{Massimiliano Maria \surname{Riva}}
\author{Filippo \surname{Vernizzi}}
\author{Leong Khim \surname{Wong}}
\affiliation{Universit\'{e} Paris-Saclay, CNRS, CEA, Institut de Physique Th\'{e}orique, 91191 Gif-sur-Yvette, France}

\begin{abstract}
We present a novel calculation of the four-momentum that is radiated into gravitational waves during the scattering of two arbitrarily spinning bodies. Our result, which is accurate to leading order in~$G$, to quadratic order in the spins, and to all orders in the velocity, is derived by using a Routhian-based worldline effective field theory formalism in concert with a battery of analytic techniques for evaluating loop integrals. While nonspinning binaries radiate momentum only along the direction of their relative velocity, we show that the inclusion of spins generically allows for momentum loss in all three spatial directions. We also verify that our expression for the radiated energy agrees with the overlapping terms from state-of-the-art calculations in post-Newtonian theory.
\end{abstract}

\maketitle


\section{Introduction}

The burgeoning field of gravitational-wave astronomy \cite{LIGOScientific:2016aoc, LIGOScientific:2017vwq} will offer new opportunities to explore questions in fundamental physics, test the nature of strong-field gravity, and constrain various binary formation and evolution channels~\cite{Berti:2015itd, LIGOScientific:2020tif, Barausse:2020rsu, LISACosmologyWorkingGroup:2022jok, LISA:2022kgy, Barausse:2012fy, Callister:2020arv}.~As binary systems with spinning black holes constitute one of the primary sources of gravitational waves, modeling precisely how spin influences a binary's inspiral is essential for making robust detections and performing accurate parameter estimation studies~\cite{Vitale:2014mka, LIGOScientific:2016wkq, LIGOScientific:2020stg}.

In the traditional approach to the two-body problem, one makes the so-called post-Newtonian (PN) expansion \cite{Blanchet:2013haa}: the equations of motion for the binary and the gravitational field are solved order by order simultaneously in powers of $G$ and~$v^2$; respectively, the gravitational constant and the square of the relative velocity between the two bodies. Since the two parameters are related by the virial theorem, this perturbative scheme is ideally suited to the study of bound orbits.

An alternative approach, which lends itself more naturally to the study of unbounded orbits (i.e., scattering encounters), is the post-Minkowskian~(PM) approximation \cite{Damour:2016gwp}. Here, one also expands in powers of~$G$, but keeps $v$ fully~relativistic. While the study of unbounded orbits may, at face value, seem far removed from the coalescing binaries that gravitational-wave detectors observe, quantities computed in one scenario can be linked to the other via, e.g., analytic continuation \cite{Kalin:2019rwq, Kalin:2019inp, Cho:2021arx, Bini:2020hmy}. Alternatively, PM calculations could also be used as inputs to improve the accuracy of the effective-one-body approach \cite{Damour:2016gwp, Damour:2017zjx, Bini:2017xzy, Bini:2018ywr, Damour:2019lcq, Antonelli:2019ytb, Khalil:2022ylj}---a popular semianalytic method for constructing waveform templates.

In recent years, rapid advancements in the PM program have been driven by the scattering-amplitudes community \cite{Neill:2013wsa, Bjerrum-Bohr:2013bxa, Luna:2017dtq, Bjerrum-Bohr:2018xdl, Kosower:2018adc, Cheung:2018wkq, Bern:2019nnu, Bern:2019crd, Cristofoli:2019neg, Cristofoli:2020uzm, Cheung:2020gyp, Bjerrum-Bohr:2021vuf, Bjerrum-Bohr:2021din, Damgaard:2021ipf, Brandhuber:2021eyq, Buonanno:2022pgc}, who (at present) have pushed out calculations in the conservative sector up to 4PM; i.e., up to $O(G^4)$ \cite{Bern:2021dqo, Bern:2021yeh}. Analogous results were also obtained independently through various worldline effective field theory (EFT) approaches \cite{Kalin:2020mvi, Kalin:2020fhe, Loebbert:2020aos, Mogull:2020sak, Dlapa:2021npj, Dlapa:2021vgp}. These results were later extended to include tidal deformation \cite{Bern:2020uwk, Cheung:2020sdj, AccettulliHuber:2020oou, Haddad:2020que, Aoude:2020onz, Bini:2020flp, Kalin:2020lmz, Cheung:2020gbf} and spin effects~\cite{Arkani-Hamed:2017jhn, Chung:2018kqs, Vines:2017hyw, Vines:2018gqi, Bern:2020buy, Guevara:2018wpp, Maybee:2019jus, delaCruz:2020bbn, Liu:2021zxr, Jakobsen:2021zvh, Jakobsen:2022fcj, Alessio:2022kwv, FebresCordero:2022jts}.

Developments in the radiative sector are more recent. The four-momentum emitted into gravitational waves by a nonspinning binary was first computed at leading (3PM) order in Refs.~\cite{Herrmann:2021lqe, Herrmann:2021tct} via the ``KMOC''~approach \cite{Kosower:2018adc}, independently in Ref.~\cite{DiVecchia:2021bdo} via the eikonal approach, and then in Ref.~\cite{Riva:2021vnj} via the worldline EFT approach. Tidal contributions were later included in Ref.~\cite{Mougiakakos:2022sic}. (See also Refs.~\cite{DiVecchia:2019myk, DiVecchia:2019kta, Bern:2020gjj, DiVecchia:2020ymx, AccettulliHuber:2020dal, Damour:2020tta, DiVecchia:2021ndb, Bini:2021gat, Mougiakakos:2021ckm, Jakobsen:2021smu, Jakobsen:2021lvp, Jakobsen:2022fcj, Bini:2021gat, Bini:2021qvf, Manohar:2022dea, DiVecchia:2022owy, DiVecchia:2022nna, Alessio:2022kwv} for related works on radiative effects.) Notably absent from the literature, however, is the inclusion of spins in the radiated observables~at~3PM.

To be precise, the outgoing waveform from a spinning binary has been computed up to 2PM in Ref.~\cite{Jakobsen:2021lvp}. Using this to compute the radiated four-momentum at 3PM is challenging, however, because of the multiscale nature of the resulting integrals, which have so far proven to be intractable unless one also performs a low-velocity expansion~\cite{Mougiakakos:2021ckm, Jakobsen:2021smu, Jakobsen:2021lvp}. Fortunately, this is not the only option, and indeed our goal in this paper is to compute the four-momentum radiated at 3PM up to quadratic order in the spins and to all orders in~the~velocity.

We bypass the aforementioned complications with the waveform by formulating the problem as an integral of the outgoing graviton momentum over phase space, weighted by (what is essentially) the square of the binary's (pseudo) stress-energy tensor.~The latter we construct by using the worldline EFT formalism, while the loop integrals that arise are computed by appropriating powerful techniques from high-energy physics~\cite{Parra-Martinez:2020dzs}; namely, reverse unitarity \cite{Anastasiou:2002yz, Anastasiou:2002qz, Anastasiou:2003yy, Anastasiou:2015yha}, a reduction to master integrals via integration by parts \cite{Tkachov:1981wb, Chetyrkin:1981qh, Smirnov:2012gma}, and differential equation methods \cite{Kotikov:1990kg, Kotikov:1991pm, Bern:1992em, Gehrmann:1999as, Henn:2013pwa, Caron-Huot:2014lda}, as previously used in Refs.~\cite{Herrmann:2021lqe, Herrmann:2021tct, DiVecchia:2021bdo, Riva:2021vnj, Mougiakakos:2022sic} for the nonspinning case.

These techniques are described in more detail in Sec.~\ref{sec:methods}. In Sec.~\ref{sec:result}, we discuss the key features of our main result, but owing to its length, we present the full expression only in the Supplemental Material~\cite{SuppMaterial}, a computer-readable version of which is available in the ancillary files attached to the arXiv submission of this paper. Also included in the Supplemental Material are explicit expressions for some of the intermediate quantities that we calculate, like the stress-energy tensor. We conclude in~Sec.~\ref{sec:conclusion}.
\looseness=-1

\section{Methods}
\label{sec:methods}

\subsection{Worldline effective field theory}
Consider the case of two spinning bodies approaching one another from infinity, and suppose that their distance of closest approach remains much larger than their individual radii. In this scenario, the details of their scattering encounter are well described by an EFT in which the two bodies are treated as point particles traveling along the worldlines of their respective centers of energy. Their dynamics are conveniently described by a Routhian~\cite{Liu:2021zxr, Porto:2008tb, Porto:2008jj, Yee:1993ya}, which for each body of mass~$m$~reads
\begin{align}
	\mathcal{R}
	=&
	- \frac{1}{2}
	\bigg( 
		m g_{\mu\nu\,} \U^\mu \U^{\mathstrut\nu}
		+
		\omega_\mu^{ab} \Sp_{ab} \U^\mu
		\nonumber\\&
		-
		\frac{1}{m} \U_a \, \U^e R_{ebcd} \Sp^{ab} \Sp^{cd}
		-
		\frac{1}{m} C_{E} E_{ab} \Sp^a{}_c \Sp{}^{cb}
	\bigg).
\label{eq:Routhian}
\end{align}

The translational degrees of freedom (d.o.f.s) of this body are encoded in four worldline coordinates~$x^\mu(\tau)$, which chart the integral curve of the four-velocity ${\U^\mu \equiv \dx x^\mu/\dx\tau}$. Only three of these are needed to specify the position of the body uniquely, however, and so we remove the remaining unphysical d.o.f.~by imposing the constraint ${\U_\mu \U^\mu = 1}$~\cite{Kalin:2020mvi, Liu:2021zxr}. Meanwhile, the body's rotational d.o.f.s are encoded in the antisymmetric spin tensor~$\Sp_{ab}$, which notably is defined in a locally flat frame with coordinates~$\{y^a\}$. Tensors defined in the general coordinate frame $\{x^\mu\}$ are transformed into the former by way of the vielbein ${e_\mu^a \equiv \partial y^a/\partial x^\mu}$ (e.g., ${\U^a \equiv e^a_\mu \U^\mu}$), which also defines for us the spin connection
${\omega_\mu^{ab} \coloneq g^{\rho\sigma} \! e^b_\sigma \nabla_{\!\mu}^{\vphantom{a}} e^a_\rho}$.
Only three of the six components in~$\Sp_{ab}$ are needed to specify the spin of the body uniquely; hence, the other three d.o.f.s are to be removed by imposing a spin supplementary condition~(SSC) \cite{Hanson:1974qy, Pryce:1948pf, Newton:1949ssc, Tulczyjew:1959ssc}.~Given the relativistic nature of the problem, the covariant~SSC, ${ \,\U^a\Sp_{ab} = 0 + O(\Sp^3)}$~\cite{Tulczyjew:1959ssc}, proves to be the most convenient~choice.

The first three terms in Eq.~\eqref{eq:Routhian} are universal, in the sense that they apply to any body with a mass monopole and spin dipole. Higher-order multipole moments, however, are sensitive to the body's internal structure, and this is why the final term in Eq.~\eqref{eq:Routhian}, which describes the self-induced quadrupole moment of the rotating body (${E_{\mu\nu} \equiv R_{\mu\rho\nu\sigma} \U^\rho \U^\sigma}$), is accompanied by the Wilson coefficient~$C_E$. Kerr black holes have ${C_E = 1}$~\cite{Porto:2008jj}, although this value can be larger for objects like neutron stars~\cite{Laarakkers:1997hb, Chia:2020psj}. Infinitely many more terms can be appended to Eq.~\eqref{eq:Routhian} should we wish to include even higher-order multipoles~\cite{Levi:2015msa}, or other finite-size effects like tidal deformations~\cite{Goldberger:2004jt, Mougiakakos:2022sic}, but these all come with higher powers of either the curvature tensors or the spin, and so are irrelevant~to~our~purposes~here.

As the Routhian behaves like a Lagrangian from the point of view of the translational d.o.f.s, but like a Hamiltonian with regards to the rotational d.o.f.s, the equations of motion for each body follow from a mixture of Euler-Lagrange and Hamilton equations \cite{LandauLifshitz}; namely,
\begin{equation}
	\frac{\delta}{\delta x^\mu} \int \dx\tau\, \mathcal{R} = 0
	\quad\text{and}\quad
	\frac{\dx}{\dx\tau}\Sp^{ab} = \{ \Sp^{ab}, \mathcal{R} \}.
\label{eq:worldline_eom}
\end{equation}
The only nontrivial Poisson bracket we require is~\cite{Porto:2008tb}
\begin{equation}
	\{ \Sp^{ab}, \Sp^{cd} \}
	=
	\eta^{ac} \Sp^{bd} + \eta^{bd} \Sp^{ac} - \eta^{ad} \Sp^{bc} - \eta^{bc} \Sp^{ad},
\end{equation}
where $\eta^{ab}$ is the Minkowski metric with a mostly minus signature. Note that the aforementioned constraints on $\U^\mu$ and $\Sp_{ab}$ should be imposed only after all functional derivatives and Poisson brackets have been evaluated.

To fully specify our EFT, we must also endow the metric~$g_{\mu\nu}$ with its own dynamics; hence, we take the full effective action to be
${ S = S_\text{EH} + S_\text{FP} + \sum_{A=1}^2\int\dx\tau_A \mathcal{R}_A }$,
where $S_\text{EH}$ is the Einstein-Hilbert action, $S_\text{FP}$~is a Faddeev-Popov term that enforces the de~Donder gauge, and the label ${A \in \{1,2\}}$ is used to distinguish between the binary's two~constituents.

\subsection{Stress-energy tensor}

As a precursor to computing the radiated four-momentum, we first determine the stress-energy tensor $T^{\mu\nu}$ for the binary~as a whole. This object is sourced by the multipolar moments of the two bodies, as well as by the energy stored in nonlinear interactions of the gravitational field, and can be obtained perturbatively from our EFT with the help of Feynman diagrams once we expand the action in powers of ${ \kappa h_{\mu\nu} \coloneq g_{\mu\nu} - \eta_{\mu\nu}}$, with~${\kappa \equiv \sqrt{32\pi G}}$. Crucially,~since
\begin{equation}
	e_\mu^a = \eta^{a\nu}
	\bigg(
		\eta_{\mu\nu}
		+
		\frac{1}{2} \kappa h_{\mu\nu}
		-
		\frac{1}{8} \kappa^2 h_{\mu\rho} h^\rho{}_\nu
		+ 
		O(\kappa^3 h^3)
	\bigg)
\end{equation}
in this expansion, Greek and Latin indices are now indistinct. The spin tensors are, nonetheless, still defined in their respective locally flat~frames~\cite{Liu:2021zxr}.

\begin{figure}
\centering
\includegraphics[width=\columnwidth]{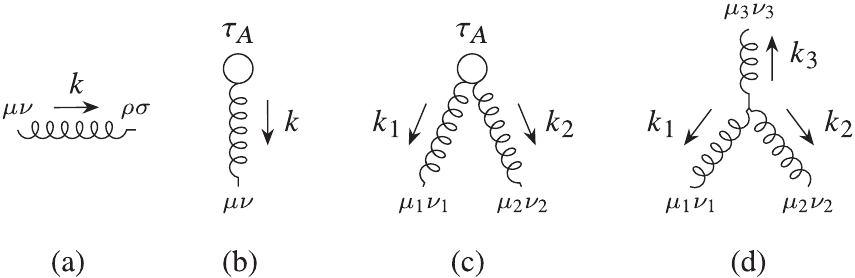}
\caption{Feynman rules relevant to our calculation.}
\label{fig:feynman_rules}
\end{figure}

Only the Feynman rules shown in Fig.~\ref{fig:feynman_rules} are required at the order in~$G$ to which we are working. In $d$-dimensional momentum space (we work in $d$~dimensions for added generality), the graviton propagator in Fig.~\ref{fig:feynman_rules}(a) is $i P_{\mu\nu\rho\sigma}/k^2$, with ${ P_{\mu\nu\rho\sigma} = \eta_{\mu(\rho} \eta_{\sigma)\nu} + \eta_{\mu\nu}\eta_{\rho\sigma}/(d-2) }$ (an~$i\epsilon$ prescription is unnecessary at this order because the loop integrals never hit the poles at~${k^2=0}$~\cite{Mougiakakos:2021ckm}), while the worldline vertex for single-graviton emission is
\begin{align}
	& \text{Fig.~\ref*{fig:feynman_rules}(b)}
	=
	-\frac{1}{2} i\kappa
	\!\int \dx\tau_A\, e^{ik\cdot x_A}
	\bigg[
	m_{A\,} \U_A{}^{\mu\,} \U_A{}^{\nu}
	+
	i k_\rho \Sp_A{}^{\rho(\mu} \U_A{}^{\nu)}
	\nonumber\allowdisplaybreaks\\&\;\;
	+
	\frac{1}{m_A} k_\rho k_\sigma \U_{A\,\alpha}
	\big(
		\U_A{}^{(\mu} \Sp_A{}^{\nu)\rho} \Sp_A{}^{\sigma\alpha}
		+
		\U_A{}^\rho \Sp_A{}^{\sigma(\mu} \Sp_A{}^{\nu)\alpha}
	\big)
	\nonumber\allowdisplaybreaks\\&\;\;
	+
	\frac{1}{2m_A} C_{E_A} k_\rho k_\sigma
	\big(
		\Sp_A{}^{\rho\alpha}\Sp_A{}^{\sigma}{}_\alpha\, \U_A{}^\mu \U_A{}^\nu
		\nonumber\allowdisplaybreaks\\&\;\;
		+
		2 \,\U_A{}^\rho \Sp_A{}^{\sigma\alpha} \Sp_{A\,\alpha}{}^{(\mu} \U_A{}^{\nu)}
		+
		\Sp_A{}^{(\mu}{}_\alpha \Sp_A{}^{\nu)\alpha} \U_A{}^\rho \U_A{}^\sigma 
	\big)
	\bigg].
\label{eq:feynman_rule_wv1}
\end{align}
Expressions for the two remaining vertices, which are much lengthier, are presented in the Supplemental Material~\cite{SuppMaterial}.

In addition to making the weak-field expansion above, we must also expand the body variables ${ \X_A \equiv (x_A, \U_A, \Sp_A) }$ about their initial straight-line trajectories in order to achieve manifest power counting in~$G$. We therefore write~\cite{Kalin:2020mvi, Liu:2021zxr}
\begin{equation}
	\X_A(\tau_A) = \overline{\X}_A(\tau_A)
	+ \sum_{n=1}^\infty \delta^{(n)}\X_A(\tau_A),
\end{equation}
where $\delta^{(n)}\X_A$ is the $O(G^n)$ deflection away from the initial trajectory $\overline{\X}_A$ due to the gravitational pull of the other body. The 1PM deflections~$\delta^{(1)}X_A$, which we will need later in our calculation, were previously computed using Eq.~\eqref{eq:worldline_eom} in~Refs.~\cite{Kalin:2020mvi, Liu:2021zxr}. (They are reproduced in the Supplemental Material~\cite{SuppMaterial} for completeness.) As for~$\overline{\X}_A$, we~write
\begin{equation}
	\overline{x}{}^\mu_A = b_A^\mu + u_A^\mu \tau_A^{\mathstrut},
	\quad
	\overline{\U}{}^\mu_A = u_A^\mu,
	\;\;\text{and}\;\;
	\overline{\Sp}{}_A^{\mu\nu} = m_A^{\mathstrut} s_A^{\mu\nu},
\end{equation}
where the constant vectors $u_A^\mu$ and $b_A^\mu$ are the initial velocity and orthogonal displacement of the $A$th body, respectively, while the constant tensor $s_A^{\mu\nu}$ describes its initial spin per unit mass. Note that ${s_A^{\mu\nu} u_{A\,\nu} = 0}$ as per the covariant~SSC, while the impact parameter ${b^\mu_{\mathstrut} \coloneq b_1^\mu - b_2^\mu}$ satisfies ${b \cdot u_A = 0}$~\cite{Kalin:2020mvi}.

\begin{figure}
\centering
\includegraphics[width=\columnwidth]{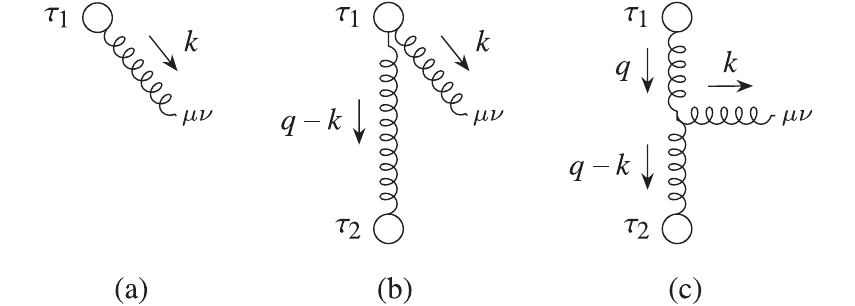}
\caption{Feynman diagrams contributing to the stress-energy tensor up to next-to-leading order in~$G$. While not drawn explicitly, our calculation includes the mirror inverses of (a)~and~(b), which are obtained by interchanging the body labels~${1 \leftrightarrow 2}$ and redefining the loop momentum ${q \mapsto k-q}$.}
\label{fig:feynman_diagrams}
\end{figure}

We now use these rules to compute the (tree-level) expectation value ${\avg{h_{\mu\nu}(k)} \equiv \kappa P_{\mu\nu\rho\sigma} T^{\rho\sigma}(k)/(2k^2)}$, from which the (classical) stress-energy tensor~$T^{\mu\nu}$ may be extracted. At~leading order in~$G$, only the diagram in Fig.~\ref{fig:feynman_diagrams}(a), with $\X_A$ replaced by~$\overline{\X}_A$, contributes. The result is
\begin{gather}
	T^{\mu\nu}_\text{LO}(k)
	=
	\sum_{A=1}^2 
	\deltabar(k\cdot u_A) m_A e^{ik \cdot b_A}
	\Big[
		u_A{}^\mu u_A{}^{\nu\mathstrut}
		+
		i k^\rho s_{A\,\rho}{}^{(\mu} u_A{}^{\nu)}
		\nonumber\\
		-
		\frac{1}{2} C_{E_A} 
		(k_\rho s_A{}^{\rho\sigma}{s}_{A\,\sigma}{}^\alpha k_\alpha)
		u_A{}^\mu u_A{}^\nu 
	\Big],
\label{eq:T_LO}
\end{gather}
where the delta function ${\deltabar(x) \equiv 2\pi \delta(x)}$ comes from having performed the integral over~$\tau_A$ in Eq.~\eqref{eq:feynman_rule_wv1}.

All three diagrams in Fig.~\ref{fig:feynman_diagrams} contribute at next-to-leading order in~$G$. From Fig.~\ref{fig:feynman_diagrams}(a), we extract the $O(G)$ part of the diagram by expanding $X_A$ up to~1PM, whereas for Figs.~\ref{fig:feynman_diagrams}(b) and~\ref{fig:feynman_diagrams}(c), it suffices to replace $\X_A$ by~$\overline{\X}_A$. The total result~is
\begin{equation}
	T^{\mu\nu}_\text{NLO}(k)
	=
	\frac{\kappa^2 M^2\nu}{4} \!\int_q 
	\frac{ \Delta_{12}(q,k) }{q^2 (k-q)^2} t^{\mu\nu}(q,k) \,e^{iq\cdot b} e^{ik\cdot b_2},
\label{eq:T_NLO}
\end{equation}
where ${M = m_1 + m_2}$ is the binary's total mass, ${\nu = m_1 m_2/M^2}$ is its symmetric mass ratio,
${\int_q \!\equiv\! \int \dx^dq/(2\pi)^d}$,
and
$\Delta_{12}(q,k) \coloneq\,\deltabar(q \cdot u_1) \,\deltabar((k - q) \cdot u_2) $.
An explicit expression for the object~$t^{\mu\nu}$, accurate to $O(s^2)$, is presented in the \mbox{Supplemental Material~\cite{SuppMaterial}}.

\subsection{Loop integrals}
Given the above, we may now compute the radiated four-momentum via the definition~\cite{Goldberger:2016iau}
 \begin{equation}
	P^\mu_\text{rad}
	=
	\frac{\kappa^2}{4} \! \int_k \deltabar_+(k^2) k^\mu 
	\big[
		T^{\alpha\nu}(k) P_{\alpha\nu\rho\sigma} T^{*\rho\sigma}(k)
	\big],
\label{eq:P_rad_def}
\end{equation}
where ${\deltabar_+(k^2) \, \dx^dk/(2\pi)^d}$ is the Lorentz-invariant phase-space measure for the emission of on shell gravitons. Observe~that because the delta functions in Eq.~\eqref{eq:T_LO} have compact support away from ${k^2=0}$, $T^{\mu\nu}_\text{LO}$ is nonradiative and so does not contribute to~$P^\mu_\text{rad}$. It therefore suffices to substitute Eq.~\eqref{eq:T_NLO} into Eq.~\eqref{eq:P_rad_def} when working to leading order in~$G$. We~then~find
\begin{align}
	P^\mu_\text{rad}
	=\:&
	\frac{\kappa^6 M^4\nu^2}{64}\int_{k,q_1,q_2}\!\!
	\frac{ t^{\alpha\nu}(q_1, k) \, P_{\alpha\nu\rho\sigma} \, t^{*\rho\sigma}(q_2,k) }{q_1^2 q_2^2 (k-q_1)^2(k-q_2)^2} k^\mu
	\nonumber\\&\times
	e^{i(q_1-q_2)\cdot b}\,
	\deltabar_+(k^2) \, \Delta_{12}(q_1, k) \Delta_{12}(q_2,k).
\end{align}

Next, we define new momentum variables ${q = q_1 - q_2}$, ${\ell_1 = -q_2}$, and ${\ell_2 = q_1-k}$~\cite{Riva:2021vnj} in order to write
\begin{equation}
	P^\mu_\text{rad}
	=
	\frac{\kappa^6 M^4\nu^2}{64} \!
	\int_q \deltabar(q\cdot{u}_1)\deltabar(q\cdot{u}_2) \, e^{iq\cdot b} Q^\mu(q).
\label{eq:P_rad_Q}	
\end{equation}
The radiated four-momentum may thus be viewed as the inverse Fourier transform of some object $Q^\mu$, which is expressible as a sum of terms in which $q^\mu$, $u_A^\mu$, and $s_A^{\mu\nu}$ are contracted amongst themselves and with the two-loop integrals
\begin{equation}
	G^{\mu_1 \cdots\, \mu_i \nu_1 \cdots\, \nu_j}_{n_1 \cdots\, n_9}
	\coloneq
	\int_{\ell_1, \ell_2}
	\frac{%
		\ell_1^{\mathstrut\mu_1} \cdots\, \ell_1^{\mathstrut\mu_i} \,
		\ell_2^{\mathstrut\nu_1} \cdots\, \ell_2^{\mathstrut\nu_j}
	}{%
		\urho_1^{n_1} \rho_2^{n_2} \rho_3^{n_3} \urho_4^{n_4}
		\rho_5^{n_5} \rho_6^{n_6} \urho_7^{n_7} \rho_8^{n_8} \rho_9^{n_9}
	}.
\label{eq:G_integrals}
\end{equation}
Following Ref.~\cite{Parra-Martinez:2020dzs}, we define
$\rho_2 = {-2\ell_1\cdot u_2}$,
$\rho_3 = {-2\ell_2\cdot u_1}$,
${\rho_5 = \ell_1^2}$,
${\rho_6 = \ell_2^2}$,
${\rho_8 = (\ell_1 - q)^2}$, and
${\rho_9 = (\ell_2 - q)^2}$,
while the underlined variables are used to denote the presence of delta functions; i.e.,
${2/\urho{}_1 = \deltabar(\ell_1 \cdot u_1) }$,
${2/\urho{}_4 = \deltabar(\ell_2 \cdot u_2) }$, and
${1/\urho{}_7 = \deltabar_+(k^2) }$
with ${k \equiv q-\ell_1-\ell_2}$.
All of the integrals in $Q^\mu$ have $n_1 = n_4 = {n_7 = 1}$. Reverse unitarity then allows us to treat these delta functions as cut propagators~\cite{Anastasiou:2002yz, Anastasiou:2002qz, Anastasiou:2003yy, Anastasiou:2015yha}.

The tensor-valued nature of these integrals make them cumbersome to evaluate as is, but fortunately they can all be reduced to scalar-valued ones via a suitable basis decomposition~\cite{Jakobsen:2021lvp}. Specifically, we expand each of the loop momenta $\ell_A^\mu$ (${A \in \{1,2\}}$) in the numerator as
\begin{equation}
	\ell_A^\mu 
	=
	(\ell_A \cdot u_1)\, \dualu_1{}^\mu
	+
	(\ell_A \cdot {u}_2)\, \dualu_2{}^\mu
	+
	\frac{(\ell_A \cdot q)}{q^2} q^\mu_{\mathstrut} + \ell_{A\bot}^\mu,
\label{eq:ell_decomposition}
\end{equation}
where ${\dualu_1^\mu \coloneq (u_1^\mu - \gamma u_2^\mu)/(1-\gamma^2)}$ and
${\dualu_2^\mu \!\coloneq \dualu_1^\mu |_{1\leftrightarrow 2} }$
are the dual vectors to the two initial velocities (${\dualu_A \cdot u_B = \delta_{AB}}$), ${\gamma \equiv u_1 \cdot u_2}$ is the Lorentz factor for the relative velocity~$v$, and $\ell_{A\bot}^\mu$ is the part of $\ell_A^\mu$ that is orthogonal to $u_1$, $u_2$, and~$q$ [the~delta functions in Eq.~\eqref{eq:P_rad_Q} guarantee that ${q \cdot u_A = 0}$]. The~three inner products in Eq.~\eqref{eq:ell_decomposition} are then easily rewritten in terms of the variables $\rho_i$ and~$q^2$ only.

As for~$\ell_{A\bot}^\mu$, the fact that the denominator of Eq.~\eqref{eq:G_integrals} is invariant under $(\ell_{1\bot}, \ell_{2\bot}) \mapsto -(\ell_{1\bot},\ell_{2\bot})$ implies that any term in the numerator with $i$ powers of $\ell_{1\bot}$ and $j$ powers of $\ell_{2\bot}$ will integrate to zero if ${i + j}$ is odd. If instead ${i+j=2}$, then rotational invariance on the hypersurface orthogonal to $u_1$, $u_2$, and $q$ allows us to replace
\begin{equation}
	\ell_{A\bot}^{\mathstrut\mu} \ell_{B\bot}^{\mathstrut\nu}
	\mapsto
	\frac{(\ell_A^{\mathstrut\rho} \bot_{\rho\sigma} \ell_B^{\mathstrut\sigma})}{d-3}
	\bot^{\mu\nu}	
\end{equation}
under the integral, where the metric on this hypersurface is
${
	\bot^{\mu\nu}
	=
	\eta^{\mu\nu} - \dualu_1{}^{\mu} u_1{}^{\nu} - \dualu_2{}^{\mu} u_2{}^{\nu}
	- q^\mu q^\nu/q^2
}$,
and note that the inner product
${(\ell_A^{\mathstrut\rho} \bot_{\rho\sigma} \ell_B^{\mathstrut\sigma})}$
is easily rewritten solely in terms of the variables $\rho_i$, $q^2$,~and~$\gamma$.
Analogous replacement rules can be derived for the ${i+j=4}$ case by positing the ansatz
$
	\ell_{A\bot}^{\mathstrut\mu} \ell_{B\bot}^{\mathstrut\nu}
	\ell_{C\bot}^{\mathstrut\rho} \ell_{D\bot}^{\mathstrut\sigma}
	\mapsto {c_1 \bot^{\mu\nu}\bot^{\rho\sigma}}
	+ {c_2 \bot^{\mu\rho}\bot^{\nu\sigma}}
	+ {c_3 \bot^{\mu\sigma}\bot^{\rho\nu}}	
$
and then solving for the coefficients ${ \{ c_1, c_2, c_3 \} }$ by taking appropriate contractions. The same can be done for all ${i+j \in 2\mathbb Z}$, although in practice we encounter only integrals~with~${i+j \leq 5}$.

The object $Q^\mu$ is now a sum of terms in which different combinations of $q^\mu$, $u_A^\mu$, and $s_A^{\mu\nu}$ are contracted with one another and multiplied by one of the scalar-valued integrals~$G_{n_1 \cdots\, n_9}$. At this stage, $3100$ different scalar integrals enter into~$Q^\mu$, but not all of them are independent. After using the \emph{LiteRed} software package \cite{Lee:2012cn, Lee:2013mka} to identify nontrivial integration by parts relations between the different integrals \cite{Tkachov:1981wb, Chetyrkin:1981qh, Smirnov:2012gma}, we find that they reduce to a set of only four master integrals; the same four as in Eqs.~(4.13)--(4.16) of Ref.~\cite{Riva:2021vnj}. These are solved via differential equation methods~\cite{Kotikov:1990kg, Kotikov:1991pm, Bern:1992em, Gehrmann:1999as, Henn:2013pwa, Caron-Huot:2014lda}; see Ref.~\cite{Riva:2021vnj}~for~details.

All that remains is to compute the Fourier transform in Eq.~\eqref{eq:P_rad_Q}. We do so by using another family of master~integrals,
\begin{equation}
	I^{\mu_1\cdots\,\mu_j}_n
	\coloneq
	\int_q \deltabar(q \cdot u_1) \deltabar(q \cdot u_2)
 	\frac{q^{\mu_1} \cdots q^{\mu_j}}{(-q^2)^n} e^{i q \cdot b},
\end{equation}
whose scalar-valued member evaluates~to
\begin{equation}
		I_n
		=
	 	\frac{\Gamma(d/2-n-1)}{\Gamma(n)}
		\frac{ (-b_\mu \bot_{12}^{\mu\nu} b_\nu )^{n+1-d/2} }%
			{ 4^n \pi^{(d-2)/2}(\gamma^2 - 1)^{1/2} },
\end{equation}
with ${\bot_{12}^{\mu\nu} = \eta^{\mu\nu}_{\mathstrut} - \check{u}_1^{\mathstrut\mu} u_1^{\mathstrut\nu}-\check{u}_2^{\mathstrut\mu} u_2^{\mathstrut\nu}}$~\cite{Jakobsen:2022fcj}.~Its tensor-valued~cousins follow from differentiation; e.g., ${I^{\mu}_n = -i\partial I_n /\partial b_\mu}$.

\section{Results}
\label{sec:result}

\subsection{Radiated four-momentum}
Now specializing to four dimensions, it becomes convenient to decompose our final result into components along the basis vectors $\{ u_1, u_2, \bhat, \lhat \}$, where ${\bhat^\mu \coloneq b^\mu/b}$ ${( b\equiv \sqrt{-b_\mu b^\mu} )}$ and
${ \lhat_\mu \coloneq \epsilon_{\mu\nu\rho\sigma} u_1^{\mathstrut\nu} u_2^{\mathstrut\rho} \bhat^{\mathstrut\sigma}_{\mathstrut} / \sqrt{\gamma^2-1} }$ are the unit vectors pointing along the impact parameter and orbital angular momentum, respectively. After also eliminating the spin tensors in favor of the Pauli-Lubanski spin vectors
$ {s_A^\mu \coloneq \epsilon^\mu_{\mathstrut}{}_{\nu\rho\sigma}u_A^{\mathstrut\nu} s_A^{\mathstrut\rho\sigma} / 2}$, we find that we can write
\begin{equation}
	P^\mu_\text{rad}
	=
	\frac{G^3 M^4 \pi \nu^2}{b^3}
	\big(
		\mathcal{C}_{u_1} \check{u}_1^\mu
		+
		\mathcal{C}_{u_2} \check{u}_2^\mu
		-
		\mathcal{C}_{\bhat} \bhat^\mu
		-
		\mathcal{C}_{\lhat} \lhat^\mu
	\big).
\label{eq:P_rad_final}
\end{equation}
The components $\mathcal{C}_V$, with ${V \in \{ u_1, u_2, \bhat, \lhat \}}$, are dimensionless functions of only the Lorentz factor~$\gamma$, the two Wilson coefficients~$C_{E_A}$, and the six inner products $(s_A \cdot V)/b$. (There are only six because ${s_A \cdot u_A = 0}$ by definition.)

\begin{table}
\caption{Functions of the Lorentz factor $\gamma$ appearing in Eq.~\eqref{eq:P_rad_components}.\hfill}
\label{table:f}
\bgroup
\scriptsize
\def\arraystretch{2.5}
\begin{ruledtabular}
\begin{tabular}{ll @{\hspace{2em}}}
{\small $f_\text{I}$}
&
$\displaystyle
\frac{210\gamma^6-552\gamma^5+339\gamma^4-912\gamma^3+3148\gamma^2-3336\gamma+1151}{48(\gamma^2-1)^{3/2}}$
\\&
$\displaystyle
-\frac{35\gamma^4+60\gamma^3-150\gamma^2+76\gamma-5}{8\sqrt{\gamma^2-1}}\log\!\left(\!\frac{1+\gamma}{2}\!\right)$
\\&
$\displaystyle
+\frac{70\gamma^7-165\gamma^5+112\gamma^3-33\gamma}{16(\gamma^2-1)^2}\cosh^{-1}\!\gamma$
\\[1.3em]
{\small $f_\text{II}$}
&
$\displaystyle
\frac{210\gamma^6-356\gamma^5-111\gamma^4-1627\gamma^3+5393\gamma^2-4741\gamma+1352}{16(\gamma+1)(\gamma^2-1)}$
\\&
$\displaystyle
-\frac{105\gamma^4+345\gamma^3-405\gamma^2+147\gamma-48}{8(\gamma+1)}\log\!\left(\!\frac{1+\gamma}{2}\!\right)$
\\&
$\displaystyle
+\frac{210\gamma^6-405\gamma^4+135\gamma^2}{16(\gamma^2-1)^{3/2}}\cosh^{-1}\!\gamma$
\\[1.3em]
{\small $f_\text{III}$}
&
$\displaystyle
\frac{210\gamma^6-279\gamma^5-219\gamma^4-1350\gamma^3+4732\gamma^2-4243\gamma+1245}{16(\gamma+1)(\gamma^2-1)}$
\\&
$\displaystyle
-\frac{21\gamma^4+66\gamma^3-84\gamma^2+30\gamma-9}{2(\gamma+1)}\log\!\left(\!\frac{1+\gamma}{2}\!\right)$
\\&
$\displaystyle
+\frac{42\gamma^6-81\gamma^4+27\gamma^2}{4(\gamma^2-1)^{3/2}}\cosh^{-1}\!\gamma$
\\[1.3em]
{\small $f_\text{IV}$}
&
$\displaystyle
-\frac{425\gamma^5-1215\gamma^4+2491\gamma^3-3957\gamma^2+2992\gamma-760}{16(\gamma+1)(\gamma^2-1)^2}$
\\&
$\displaystyle
-\frac{84\gamma^6+459\gamma^5-825\gamma^4-138\gamma^3+666\gamma^2-321\gamma+75}{8(\gamma+1)(\gamma^2-1)^2}\log\!\left(\!\frac{1+\gamma}{2}\!\right)$
\\&
$\displaystyle
+\frac{168\gamma^7+78\gamma^6-414\gamma^5-171\gamma^4+261\gamma^3+81\gamma^2-27\gamma}{16(\gamma+1)(\gamma^2-1)^{5/2}}\cosh^{-1}\!\gamma$
\end{tabular}
\end{ruledtabular}
\egroup
\end{table}

The fact that $P^\mu_\text{rad}$ is a polar vector strongly constrains which inner products can appear at any given order, and in which combinations. For instance, because $\mathcal{C}_{u_1}$, $\mathcal{C}_{u_2}$, and $\mathcal{C}_{\bhat}$ must all be even under parity, they can only depend on $(s_A \cdot \lhat)/b$ at linear order in the spins. Indeed, we find explicitly~that
\begin{align}
	\mathcal{C}_{u_1}
	&=
	f_\text{I}(\gamma)
	+
	\frac{1}{b}
	\big[
		(s_1 \cdot \lhat) f_\text{II}(\gamma)
		+ (s_2 \cdot \lhat) f_\text{III}(\gamma)
	\big]
	+
	O(s^2),
	\nonumber\\
	\mathcal{C}_{\lhat}
	&=
	\frac{1}{b}
	\big[ (s_1 \cdot u_2) + (s_2 \cdot u_1) \big]
	f_\text{IV}(\gamma)
	+
	O(s^2),
\label{eq:P_rad_components}
\end{align}
while ${\mathcal{C}_{\bhat} = 0 + O(s^2)}$. The remaining component $\mathcal{C}_{u_2}$ can be obtained from $\mathcal{C}_{u_1}$ by swapping the body labels~${1\leftrightarrow 2}$, since $P^\mu_\text{rad}$ must be symmetric under this interchange. As an added consequence, $\mathcal{C}_{\bhat}$ and $\mathcal{C}_{\lhat}$ must be odd and even under this~interchange,~respectively.
\looseness=-1

The four functions ${ \{ f_\text{I}, \,\dots, f_\text{IV} \} }$ in Eq.~\eqref{eq:P_rad_components} depend purely on~$\gamma$ and are presented in Table~\ref{table:f} [$f_\text{I}$, which appears at $O(s^0)$, was previously determined in Refs.~\cite{Herrmann:2021lqe, Herrmann:2021tct, DiVecchia:2021bdo, Riva:2021vnj}, but is reproduced here for completeness]. An additional 21 functions of~$\gamma$, with similar analytic structures, appear at~$O(s^2)$. These are presented in the Supplemental Material~\cite{SuppMaterial}. Notice that $\mathcal{C}_{\bhat}$ and~$\mathcal{C}_{\lhat}$ both vanish when the spins are aligned along~$\lhat$ (or, indeed, when they are zero); hence, for so-called ``aligned-spin'' configurations, for which the binary's motion is confined to a plane, we see that momentum is lost only in the direction of~the~relative~velocity.

\subsection{Consistency checks}
To validate Eq.~\eqref{eq:P_rad_final} against the existing literature, we compare results for the energy $\Delta E$ radiated in the center-of-mass frame. This is computed in our approach as ${\Delta E = (P_\text{rad} \cdot p_\text{tot})/E}$, where ${p^\mu_\text{tot} = m_1^{\mathstrut} u_1^\mu + m_2^{\mathstrut} u_2^\mu}$ and ${E^2 = p_\text{tot}^2 = M^2[1 +2\nu(\gamma-1)]}$. Since $\bhat^\mu$ and $\lhat^\mu$ are purely spatial in this frame, ${s_A \cdot \bhat}$ and ${s_A \cdot\lhat}$ are equivalent to the three-dimensional dot products ${-\,\mathbf{s}_A \cdot \hat{\vphantom{b}}\kern-1.4pt\mathbf{b}}$ and ${-\,\mathbf{s}_A \cdot \hat{\mathbf{l}}}$, respectively, and note that ${s_1 \cdot u_2 \simeq \mathbf{s}_1 \cdot \mathbf{v}}$ while ${s_2 \cdot u_1 \simeq -\,\mathbf{s}_2 \cdot \mathbf{v}}$ after expanding to first order in the relative 3-velocity~$\mathbf{v}$. Having done so, our result for $\Delta E$ agrees with that of Ref.~\cite{Jakobsen:2021lvp}, which is accurate to leading order in~$\mathbf{v}$ and to quadratic order in the spins, once we also replace $(\mathbf{b}, \mathbf{s}_A, C_{E_A}) \mapsto (-\mathbf{b}, -\mathbf{s}_A, 1-C_{E_A})$ to account for differing~conventions.

As a second consistency check, we use analytic continuation by way of the boundary-to-bound (B2B) map \cite{Kalin:2019rwq, Kalin:2019inp, Cho:2021arx} to convert our result for $\Delta E$ into the energy $\Delta E_\text{ell}$ radiated during one period of ellipticlike motion. This is accomplished in three steps. Owing to current limitations of the B2B map, we first specialize to aligned-spin configurations. Next, we must transform from the covariant SSC to the canonical (Newton-Wigner) SSC~\cite{Newton:1949ssc} for the map to work. This generally entails transforming $(\mathbf{b}, \mathbf{s}_A)$ to new canonical variables $(\mathbf{b}_\text{c}, \mathbf{s}_{A\,\text{c}})$ \cite{Vines:2017hyw, Vines:2018gqi}, but ${ \mathbf{s}_A \equiv \mathbf{s}_{A\,\text{c}} }$ in the aligned-spin case; hence, only the magnitude of the impact parameter must be transformed.~The~rule~is
\begin{equation}
	b p_\infty
	=
	b_\text{c} p_\infty  - \frac{E-M}{2E} \big[ Ea_+ - (m_1 - m_2) a_- \big],
\end{equation}
where ${p_\infty =  M^2 \nu \sqrt{\gamma^2 -1}/E}$ is the initial momentum of either body in the center-of-mass frame, ${a_\pm = (\mathbf{s}_1 \!\pm \mathbf{s}_2) \cdot \hat{\mathbf{l}} }$, and we may define ${L_\text{c} = b_\text{c} p_\infty}$ as the canonical orbital angular momentum. Finally, we obtain $\Delta E_\text{ell}$ from $\Delta E$ via~\cite{Cho:2021arx}
\begin{equation}
	\Delta E_\text{ell}(\mathcal{E}, L_\text{c}, a_{\pm})
	=
	\Delta E(\mathcal{E}, L_\text{c}, a_{\pm})
	-
	\Delta E(\mathcal{E}, - L_\text{c}, - a_{\pm}),
\end{equation}
having eliminated $\gamma$ in favor of~${\mathcal{E} \equiv (E-M)/(M\nu)}$. The left-hand side follows after analytic continuation from positive to negative values of~$\mathcal{E}$. Expanded in powers of~$\mathcal{E}$, we find that our result, which is valid in the large-angular-momentum limit~\cite{Cho:2021arx}, agrees with the overlapping terms from PN theory up to 3PN in Ref.~\cite{Cho:2021arx}, and up to 4PN~in~Refs.~\cite{Cho:2022syn, PortoPC}.

\section{Conclusion}
\label{sec:conclusion}

The worldline EFT approach has proven to be an efficient way of obtaining classical radiated observables from the scattering encounter of two compact objects, be they Schwarzschild or Kerr black holes, or even neutron stars. This work closes an important gap in the PM literature by computing the radiated four-momentum at 3PM up to quadratic order in the spins and to all orders in the velocity. Remarkably, integrating over the loop momenta required knowledge of only four master integrals---the same four as in the nonspinning case---which explains why the analytic structure of our result is similar to that of Refs.~\cite{Herrmann:2021lqe, Herrmann:2021tct, DiVecchia:2021bdo, Riva:2021vnj, Mougiakakos:2022sic}, despite the inclusion of spins. At low velocities, our radiated energy is consistent with the existing literature, including the case of the energy loss from a bound system during a single orbit, which we derived via analytic~continuation.

These results should prove invaluable for making further consistency checks in the future. We expect, for example, that our 3PM result for the radiated energy should reemerge in the tail term of the (as yet unknown) conservative potential for spinning binaries at~4PM, analogously to how the tail term \cite{Bern:2021dqo, Bern:2021yeh, Dlapa:2021npj} was found to match the radiated energy \cite{Herrmann:2021lqe, Herrmann:2021tct, DiVecchia:2021bdo, Riva:2021vnj} in the nonspinning case. In the future, it would also be interesting to reconstruct the radiated flux for bound systems (via the approach in~Ref.~\cite{Cho:2021arx}) from our calculation of the energy loss, so as to make a more direct comparison with PN~results, since it is the former that directly impacts the binary's inspiral via the balance~equation~\cite{Blanchet:2013haa}.

\acknowledgments
It is a pleasure to thank Gihyuk Cho, Gregor K\"{a}lin, and Rafael Porto for providing us with their result for the radiated energy up to~4PN. We acknowledge use of the \emph{xAct}~package~\cite{xAct} for \emph{Mathematica} in our calculations. This work was partially supported by the Centre National d'\'{E}tudes Spatiales~(CNES).\\

\emph{Note added.}---While this manuscript was undergoing peer review, we became aware of similar calculations being undertaken by Gustav Jakobsen and Gustav Mogull. We~thank them for verifying that their result for the radiated four-momentum is in agreement with ours.

\bibliography{main}
\end{document}